\documentclass{aastex631}

\received{}
\revised{}
\accepted{\today}
\submitjournal{AJ}

\shorttitle{M31 UV SNRs}
\shortauthors{Leahy, Monaghan \& Ranasinghe}
\begin{document}

\title{Discovery of 20 UV Emitting SNRs in M31 with UVIT}


\correspondingauthor{Denis Leahy}
\email{leahy@ucalgary.ca}

\author[0000-0002-4814-958X]{Denis Leahy}

\author{Christopher Monaghan}
\author{Sujith Ranasinghe}
\affiliation{Dept. Physics and Astronomy, University of Calgary, Calgary, AB, Canada;
}

\begin{abstract}
We present the first catalog of supernova remnants (SNRs) in M31 which exhibit diffuse ultraviolet (UV) emission. 
UV images of M31 were obtained by the Ultraviolet Imaging Telescope (UVIT) on the AstroSat satellite, and 
the list of SNRs was obtained from X-ray, optical and radio catalogues of SNRs in M31.
We used the UVIT images to find SNRs with diffuse emission, omitting those too contaminated with stellar emission.
20 SNRs in M31 were detected with diffuse UV emission. 
Fluxes in the UVIT F148W, F169M, F172M, N219M and N279N filters are measured for these SNRs. 
The luminosities are compared to those computed from the spectra of seven known UV-emitting SNRs in the Milky Way, the LMC, and the SMC. 
We find similar spectral shapes between the known and the M31 UV-emitting SNRs.
The spectral shapes and the diffuse nature of the emission are good evidence that the UV emissions are dominated by line emissions, like known SNRs, and the UV is associated with the SNRs.
Models are applied to the 6 SNRs with X-ray spectra. 
The main difference is that the 2 X-ray/UV SNRs are Type Ia and the 4 X-ray/non-UV SNRs are core-collapse or unknown type. 
A comparison of M31 SNRs in different wavebands shows that most  are detected optically, similar to the case for other nearby galaxies.
19 of the 20 UV-emitting SNRs are detected optically, expected because both UV and optical are from forbidden and recombination lines from shock-ionized gas.
\end{abstract}


\section{Introduction} \label{sec:intro}

A supernova remnant (SNR) is an extended (pc scale) structure in the interstellar medium excited by the shock wave from the explosive death of a star at the end of its life. 
The study of SNRs is crucial to our understanding of supernova explosions, the nature of shock waves and the structure of the interstellar medium. 
SNRs emit over a wind range of wavelengths, thus Galactic and extragalactic surveys of SNRs have been conducted at optical, radio, and X-ray wavelengths. 
There are 294 known SNRs within the Milky Way \citep{2019JApA...40...36G}. 
A small number of nearby external galaxies have had their SNRs catalogued, including the Large and Small Magellanic Clouds (LMC \& SMC), M33, NGC 300 and M31. 
E.g. there are 62 confirmed SNRs and 30 SNR candidates in the LMC, and 21 SNRs and 2 candidates in the SMC \citep{2021MNRAS.500.2336Y,2019A&A...631A.127M}. 
There are 109 SNRs and SNR candidates in M33 \citep{2000immm.proc..179D,2000ApJ...544..780P} and 44 in NGC 300 \citep{2000ApJ...544..780P}.
Optical surveys of M31 have found 156 SNRs \citep{2014ApJ...786..130L} and XMM-Newton X-ray found 26 SNRs and 21 SNR candidates \citep{2012AA...544A.144S}.
Radio emission was found for 30 SNRs in M31 by \citet{1993A&AS...98..327B}.

Galactic and extragalactic searches for SNRs have been carried out at optical, radio, and X-ray wavelengths. 
However, ultraviolet (UV) observations of SNRs are scarce. 
The difficulty in detecting UV emission is caused by the strong interstellar extinction for our Galaxy in the UV (e.g. \citealt{2021ApJS..254...38S}). 
As a result, only nearby Galactic SNRs have detected UV radiation. 
However, UV emission lines from SNRs can provide valuable information on the SNR, including shock velocities, densities, and thermal structure \citep{1997ApJ...482..881R}. 

The first major steps in analyzing the UV emission of SNRs came from the International Ultraviolet Explorer (IUE), designed for analyzing UV spectra. 
IUE data provided the groundwork for the first UV-based analyses of SNRs throughout the 1980's. 
Many studies focused on comparing theoretical models of shockwaves to nearby SNRs, such as Vela and the Cygnus Loop, by analyzing the individual line emissions in the UV spectrum \citep{1980ApJ...238..881R,1981ApJ...246..100R,1988ApJ...324..869R}.
SNRs in the Large Magellanic Cloud and the Small Magellanic Cloud, such as N49, N63, and E0102, were also analyzed using their UV spectra \citep{1980ApJ...238..601B,1992ApJ...394..158V,1980ApJ...238..601B}. 
Despite the large distances to these sources, their positions with respect to the plane of the Milky Way allows light from the LMC and SMC to pass through less of the interstellar medium, and experience less extinction than most Galactic SNRs. 
Such studies determined that the UV spectra of these remnants are dominated by line emission, with many of the same lines present in different SNRs. 
N[V], Si[IV], O[IV], He[II], and O[III] emission lines were observed in more than six intergalactic and extragalactic (LMC and SMC) SNRs, and a number of other lines were identified in more than one SNR \citep{1996ApJS..106..563F}. 

Despite the progress in UV-based SNR research, there does not yet exist a catalogue of extragalactic UV-emitting SNRs. 
As an important step in UV studies of SNRs, we conduct a search for SNRs in M31 using data from AstroSat's UVIT instrument, and generate the first catalog of UV SNRs in another galaxy. 
Section~\ref{sec:obs} below summarizes the observations and Section~\ref{sec:analysis} describes our data analysis. 
In Section~\ref{sec:cat} the catalog of UV-emitting SNRs in M31 is given, and in Section~\ref{sec:known}  the UV spectral shapes of known UV-emitting SNRs are compared to those of the M31 SNRs.  
The set of 6 SNRs with X-ray spectra are fit with SNR models to derive their physical conditions in Section~\ref{sec:SNRmodel}.
The statistics of numbers of SNRs detected in different wavebands is discussed in Section~\ref{sec:SNRnumbers} and 
 Section~\ref{sec:conclusion} summarized the results from this study.

\section{Observations} \label{sec:obs}

The observations of M31 were carried out by the Ultraviolet Imaging Telescope (UVIT) onboard the AstroSat \citep{2014SPIE.9144E..1SS}.  
UVIT is capable of observing in a variety of Far Ultraviolet (FUV) and Near Ultraviolet (NUV) bandpasses. 
The M31 survey includes data with the filters: F148W, F154W, F169M, F172M, N219M, and N279N filters, although the F154W filter was used only for one observation. 
The filter parameters, including effective area curves, can be found in \citet{2017AJ....154..128T}.
New in-orbit calibrations of UVIT were carried out by \citet{2020AJ....159..158T}. 
Data processing was carried out using CCDLab \citep{2017PASP..129k5002P,2021JApA...42...30P} to produce images with a pixel scale of $0.4168'' \times 0.4168''$ from the instrument data. 
The resulting spatial resolution, using the latest UVIT calibrations and data processing procedure, is $\simeq1$\arcsec. 

The M31 survey with UVIT \citep{2020ApJS..247...47L} consists of 19 partially-overlapping fields, each with a diameter of $\approx 28'$, covering a sky area of $\approx 3.3^{\circ} \times 1.3^{\circ}$. 
Since the survey paper \citep{2020ApJS..247...47L}, additional observations were carried out, including observation of the missing field number 8 in F148W and F169M bands to yield full coverage of the M31 survey area in the F148W filter, and partial coverage in the other filters. 
The images of these 19 fields, each with 2 to 5 filter bands, were used in the analysis carried out here.  

\section{Data Analysis} \label{sec:analysis}

\subsection{Selection of SNRs}  \label{sec:selection}

The list of SNRs and SNR candidates within M31 was obtained from existing optical, X-ray 
and radio SNR lists\footnote{\citet{2014SerAJ.189...15G} gives a catalog of 916 radio sources in M31 detected at 20cm, however the sources which are SNRs are not identified.}.
The optical SNRs were from \citet{2014ApJ...786..130L}, which contained 156 SNR candidates observed using H$\alpha$ and S[II] images. 
The X-ray SNRs were from \citet{2012AA...544A.144S} which lists 26 confirmed SNRs and 21 SNR candidates. 
These sources were first catalogued in an XMM-Newton survey of M31, which catalogued a total of 1897 X-ray sources \citep{2011AA...534A..55S}.
The radio SNRs were from  \citet{1993A&AS...98..327B} which lists 24 high confidence ($>5\sigma$)
and 6 medium confidence (3 to 5$\sigma$) detections at 1465 MHz. 
That work found 52 SNRs and candidates using narrow band imaging in [SII] and H$\alpha$ filters of the NE half of M31, then matched those with radio continuum imaging of M31.

\begin{figure}[h]
    \centering
\gridline{\fig{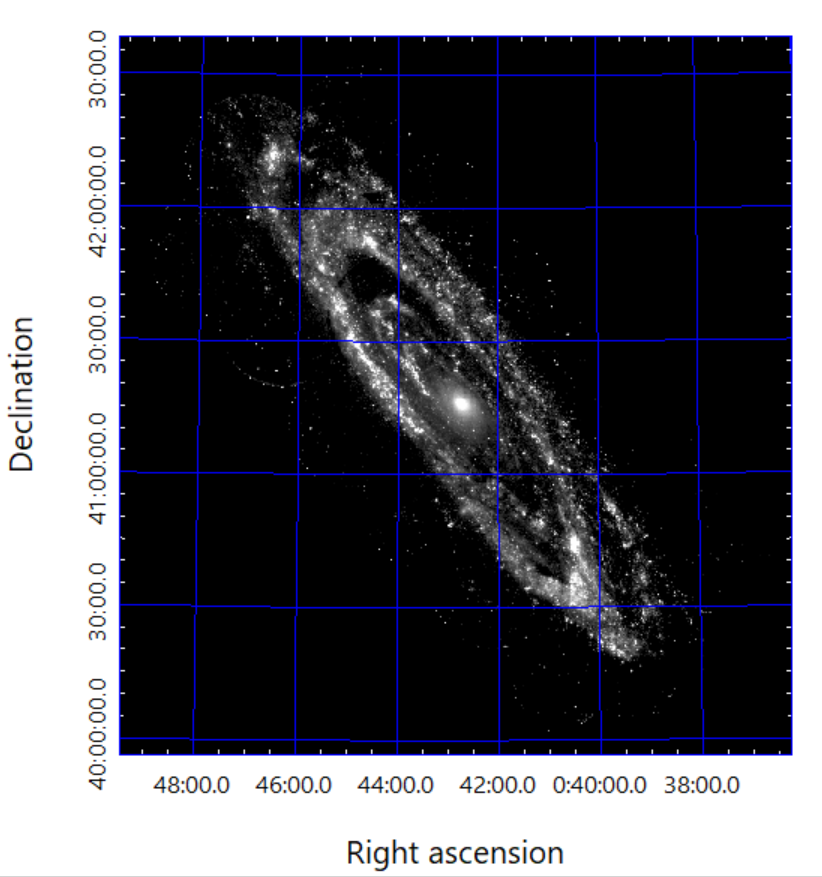}{0.49\textwidth}{(a)}
             \fig{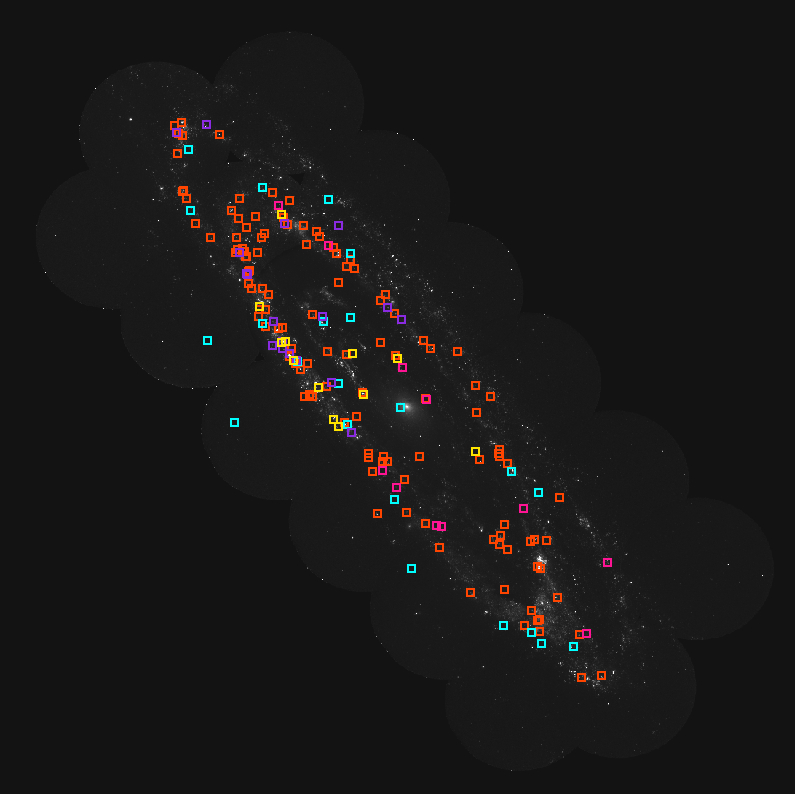}{0.51\textwidth}{(b)}
            }
    \caption{(a): the F148W mosaic of M31 (from \citealt{2020ApJS..247...47L}) but with missing field 8 added. (b): positions of the 177 SNRs from \citet{2014ApJ...786..130L} and  \citet{2012AA...544A.144S}, overlaid on the mosiac (with the mosiac made fainter to clearly see the SNRs). 
   Re squares indicate the positions of SNRs detected in optical, aqua indicates SNRs detected in X-ray,  purple indicates SNRs detected in optical and radio, pink indicates SNRs detected in optical and X-ray, 
and yellow indicates SNRs detected in optical, radio and X-ray.
SNRs identified in optical, X-ray and radio in panel (b) closely trace the positions of the spiral arms and star formation which are bright in the F148W image of M31 (panel (a)).} 
    \label{fig:m31posn}
\end{figure}

We combine the three sets of SNRs. 
The 52 optically-detected SNRs (including 30 with radio emission) from \citet{1993A&AS...98..327B} were 
included in the study of \citet{2014ApJ...786..130L}.
Four of the 30 with radio emission were re-examined by \citet{2014ApJ...786..130L} and found not to be SNRs, and one more was  found by us to be contaminated by stellar emission. 
This left 25 SNRs from \citet{1993A&AS...98..327B} that are detected in radio, which 
are included in the list in \citet{2014ApJ...786..130L}. 
Some SNRs are listed in both optical and X-ray catalogues as  
noted in the analysis of X-ray SNRs in the northern disc of the M31 
 \citep{2018AA...620A..28S}. 
Using an angular separation of $18''$, we found an additional 18 source matches between the radio and X-ray SNR catalogs. 
This left 179 unique SNRs or SNR candidates, hereafter referred to as SNRs, within M31. 
However, two of these are outside the M31 UVIT survey region, leaving 177 unique sources for analysis using UVIT data: 
119 detected only in optical, 22 detected only in X-ray, 13 detected in optical and radio, 11 detected in optical and X-ray,  and 12 detected in optical, X-ray and radio. 
The location of these 177 SNRs in M31 for the different categories is shown in Figure~\ref{fig:m31posn}(b). 

\subsection{Search for UV emission from the SNRs}  \label{sec:selection}

The locations of the 177 SNRs were used to determine within which field each object was located and thus which filters were oberved for each SNR.
Next, we carried out a set of tasks to find those SNRs which have diffuse emission not too contaminated by UV emitting stars in M31.
SNRs in UV are characterized by diffuse emission, in contrast to stars which are unresolved point sources.
UVIT has high enough spatial resolution in most cases to separate diffuse from the stellar point-source emission.

An initial inspection of the 177 SNRs was undertaken 
to remove any sources from the list without any UV emission within the optical radius of the SNR, given in \citet{2014ApJ...786..130L}. 
For the 22 X-ray-only SNRs, no optical SNR radius was available, and neither was an X-ray radius, so the position error was used instead. 
The radius of analysis around each source (whether from optical or X-ray) is henceforth referred to as the ``SNR radius''. 
Several SNRs were contained within densely packed, UV-emitting star clusters, so that no SNR emission could be distinguished from the stellar emission. 
The ``no emission'' and crowded sources were removed from the list of SNRs for analysis, leaving 126.

Further inspection of these 126 SNRs often revealed stars within the radii of the SNRs. 
Thus, we search for known stars within the SNR radius of each SNR:  
Vizier (\url{https://vizier.cds.unistra.fr/}) was used to search through stellar catalogues. 
A number of stellar catalogues were initially searched, but most were contained within the GAIA Early Data Release 3 (EDR3) catalogue \citep{BailerJones2021}. 
We found all EDR3 stars within a 15\arcsec search radius (15\arcsec was the largest optical radius of any source). 
A total of 766 stars were found in the EDR3 data (many would not have UV emission), and their coordinates were imported into CCDLAB to search for stellar contamination. 
42 sources were removed from our SNR list because all UV emission within the SNR radius was identified with catalogued GAIA stars,
leaving 84 sources in our list of SNRs. 
This process was repeated using the full data release (DR3) upon its publication on June 13th, 2022. 
5 additional stars were determined to be nearby the SNRs, but none associated with the remaining 84  with UV emission.

The 84 sources were further analyzed by searching for stellar sources from additional stellar catalogues. 
A number of catalogues were searched during this process, including a BVRI analysis of M31 objects using the McGraw Hill Telescope, a Swift/UVOT source catalogue, and a XMM-OM object survey \citep{1992A&AS...96..379M,2014Ap&SS.354...97Y,2012MNRAS.426..903P}. 
Vizier searches from these catalogues did not provide additional stellar contaminants in the 84 SNRs. 
However, the catalogue of M31 supergiants from the local group galaxy survey \citep{2016AJ....152...62M} provided additional stellar sources within a 15\arcsec search radius of the 84 SNRs. 
The red supergiant catalogue of \citet{2021AJ....161...79M} was also included in the Vizier search. 
610 supergiants from these two catalogues were found to be nearby our remaining SNRs: most of these were blue supergiants or luminous blue variables. 
This process led to a number of UV emissions to be reclassified as supergiant stars.
43 sources were removed from our SNR list, leaving 41 SNRs. 
One more source was removed using the HST M31 PHAT catalogue \citep{2014ApJS..215....9W} using PHAT stars with F275W magnitude brighter than 20.

Out of the 40 SNRs that remained, 7 had UV emission with no associations with stars. 
The other 33 consisted of either i) sources with diffuse emissions that overlapped with GAIA stars or Massey supergiants or ii) diffuse emission too dim to be reliably measured.
Stars within diffuse emission were examined to determine if they were likely to emit UV radiation: 
Larger U-B values indicated that the star emitted very little UV radiation, and would not contaminated the UV measurements. 
These 33 SNRs were analyzed in CCDLAB to determine whether or a measurement of the diffuse emission could be done. 
13 of these were isolated enough from nearby UV sources to analyze using box measurements. 
5 sources had clear indications of diffuse emission, but the region had too many overlapping stars (i.e. was confused) for a reliable flux measurement to be taken. 
The remaining 15 sources were removed from the list as there was either no clear indication of SNR emissions within the SNR radius, or the measured flux was too dim compared to the background to result in a reliable measurement.

\begin{figure}[h]
    \centering
    \includegraphics[scale=0.5]{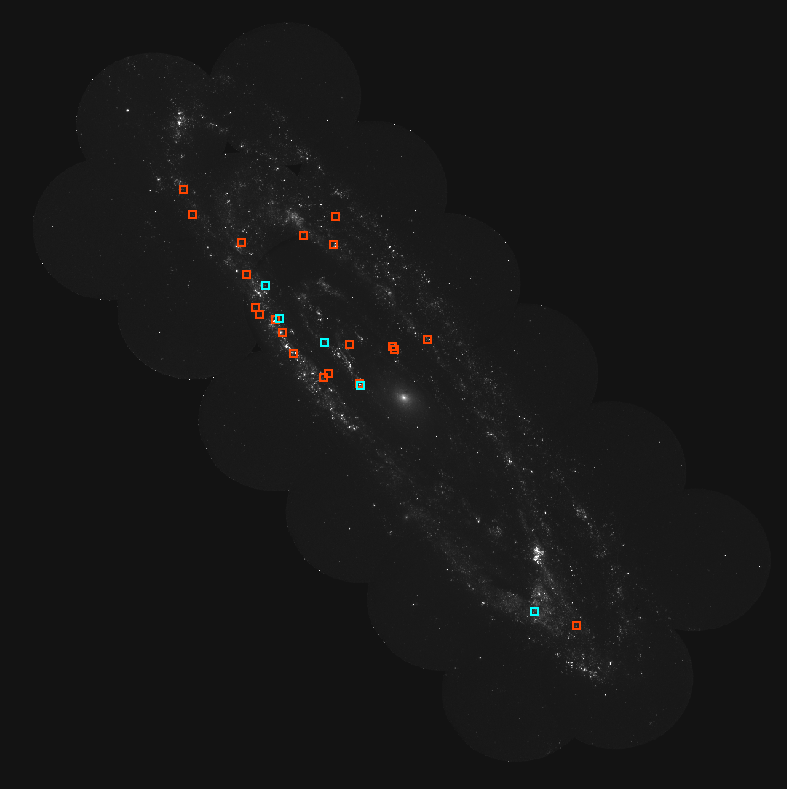}
    \caption{Positions of the 20 SNRs with detected diffuse UV emission (red squares) and of the 5 SNRs with likely, but confused, diffuse emission (blue squares), overlaid on the image of M31 in the F148W filter.} 
    \label{fig:25posn}
\end{figure}

The result of the above selection process left 25 UV emitting SNRs consisting of 7  without stellar contamination; 13 with  stellar contamination which can be separated from the diffuse emission and 5 more with likely diffuse emission but too confused with  stars for measurement. 
The locations of the 20 SNRs in M31 with clear diffuse UV emission and 5 SNRs with likely diffuse emission are shown in Figure~\ref{fig:25posn}. 
We do not consider further the 5 diffuse but confused sources for which reliable UV fluxes could not be obtained. 
Thus, the number of SNRs for which we can carry out flux measurements is 20.

Properties of the 20 SNRs with detected diffuse UV emission are shown in Table~\ref{tab:optXray}. 
Table~\ref{tab:optXray} lists each SNR's ID numbers, including their original Lee and Stiele (SPH11) IDs, their J2000 coordinates, the optical diameter and error, the likely progenitor type, and their optical and/or X-ray luminosities.

\begin{longrotatetable}
\begin{deluxetable*}{ccccccccccccc}
\tablecaption{Optical and X-ray data of the 20 UV-emitting SNRs\label{tab:optXray}} 
\tablewidth{0pt}
\tabletypesize{\scriptsize}
\tablehead{
\multicolumn{4}{l}{IDs} & \multicolumn{2}{c}{Coordinates\textsuperscript{b}}  & \multicolumn{2}{c}{Observational}   & \colhead{SNR} & \multicolumn{2}{c}{Optical Luminosity \textsuperscript{c}} & \multicolumn{2}{c}{X-ray Luminosity} \\
\multicolumn{4}{l}{ } & \multicolumn{2}{c}{(degrees)}  & \multicolumn{2}{c}{Radii (pc)}   & \colhead{Type} & \multicolumn{2}{c}{log(erg/s)} & \multicolumn{2}{c}{(erg/s)} \\
\colhead{SNR} & \colhead{Candidate\textsuperscript{a}} & \colhead{ Lee\textsuperscript{c}} & \colhead{SPH11\textsuperscript{d}} & R.A.      & DEC.    & Dia.\textsuperscript{c} & Err.\textsuperscript{e} &
\colhead{} & log(Halpha)   &log({[}SII{]}) & 0.35-2keV\textsuperscript{e} & 0.3-10keV\textsuperscript{f} 
}
\decimalcolnumbers
\startdata
1      & 5            & 6         &           & 9.9685                     & 40.4951                    & 71                    &                       & CC              & 36.88                & 36.71                   &                      &                      \\
2      & 45           & 40        &           & 10.5842                    & 41.4652                    & 45.2                  &                       & CC              & 36.74                & 36.52                   &                      &                      \\
3      & 57           & 50        & 1066      & 10.7236                    & 41.4310                    & 26.8                  & 1.89                  & Ia              & 36.49                & 36.45       & 3.80E+36             &                      \\
4      & 59           & 52        &           & 10.7326                    & 41.4402                    & 59.8                  &                       & Ia              & 36.22                & 36.17                   &                      &                      \\
5      & 75           & 67        &           & 10.8721                    & 41.3178                    & 46                    &                       & CC              & 36.28                & 36.16                   &                      &                      \\
6      & 78           & 70        & 1275      & 10.9133                    & 41.4483                    & 25.2                  & 2.12                  & Ia              & 36.6                 & 36.56      & 3.10E+36             & 1.80E+36             \\
7      & 90           & 78        &           & 10.9777                    & 41.8817                    & 37.8                  &                       & Ia              & 36.41                & 36.42                   &                      &                      \\
8      & 91           & 79        &           & 10.9838                    & 41.7853                    & 90.6                  &                       & CC              & 36.49                & 36.37                   &                      &                      \\
9      & 94           & 82        &           & 11.0045                    & 41.3515                    & 13                    &                       & CC              & 35.47                & 35.21                   &                      &                      \\
10     & 98           & 85        &           & 11.0231                    & 41.3364                    & 54.6                  &                       & CC              & 36.17                & 36.02                   &                      &                      \\
11     & 107          & 93        &           & 11.1101                    & 41.8165                    & 61.6                  &                       & CC              & 35.89                & 36.02                   &                      &                      \\
12     & 113          & 98        &           & 11.1526                    & 41.4182                    & 69.4                  &                       & CC              & 36.97                & 36.76                   &                      &                      \\
13     & 121          & 106       & 1522      & 11.1962                    & 41.4891                    & 42                    & 4.83                  & CC              & 36.59                & 36.54            & 4.40E+35   & 1.60E+35             \\
14     & 128          & 113       &           & 11.2269                    & 41.5312                    & 43                    &                       & CC              & 36.06                & 35.97                   &                      &                      \\
15     & 136          &           & 1587      & 11.2941                    & 41.5476                    &                       & 4.69                  &                 &                      &                         & 3.60E+35             &                      \\
16     & 142          & 125       &           & 11.3135                    & 41.5735                    & 69.6                  &                       & Ia              & 35.85                & 35.9                    &                      &                      \\
17     & 147          & 130       &           & 11.3517                    & 41.6839                    & 44                    &                       & CC              & 35.98                & 35.65                   &                      &                      \\
18     & 154          & 137       &           & 11.3734                    & 41.7918                    & 66.2                  &                       & Ia              & 35.96                & 36.1                    &                      &                      \\
19     & 166          & 147       &           & 11.5838                    & 41.8833                    & 89.4                  &                       & CC              & 36.49                & 36.13                   &                      &                      \\
20     & 169          & 148       &           & 11.6234                    & 41.9695                    & 49.6                  &                       & CC              & 36.21                & 36.13                   &                      &                      \\
\enddata
\tablecomments{ a) ID number from our initial list of 177 SNRs.  b) From \cite{2014ApJ...786..130L}, except for SNR 15, which is from \cite{2011AA...534A..55S}.  
c) From \cite{2014ApJ...786..130L}.  d) From \cite{2011AA...534A..55S}.
 e) From \cite{2012AA...544A.144S}.  f) From \citet{2018AA...620A..28S}.
}
\end{deluxetable*}
\end{longrotatetable}

\subsection{Measurement of SNR UV Luminosities}

From the measured fluxes and the known distance to M31, we determine the SNR UV luminosities in the different filters.
CCDLAB provides source fitting methods for photometric measurements \citep{2017PASP..129k5002P}.  
The methods include Gaussian and Moffat functional fits, a Curve-of-Growth fit (COG), and a box fit. 
For diffuse emission Gaussian and Moffat functions, designed for fitting point sources, are not suitable.
The COG fit works well to measure all counts within a given radius if there is a good background outside that radius, but does not work well for our diffuse sources usually surrounded by stellar sources. 
Thus, we use the box method to calculate the fluxes of diffuse emission and to subtract background emission while avoiding stellar emission in both source and background boxes. 

\begin{deluxetable*}{cccccccccccccc}
\tablecaption{M31 SNRs with UV emission\label{UVdata}}
\tablewidth{700pt}
\tabletypesize{\scriptsize}
\tablehead{
  & \multicolumn{2}{c}{Coordinates}                 & \multicolumn{5}{l}{UV Luminosity}  \\    
  & \multicolumn{2}{c}{(J2000)}  & \multicolumn{5}{l}{($10^{35}$erg/s)} \\
  \colhead{SNR} & \colhead{R.A.} & \colhead{DEC.} & \colhead{F148W}       & \colhead{F169M}   & \colhead{F172M}    & \colhead{N219M}   &   \colhead{N279N}      \\
 \colhead{ID} & \colhead{(hh mm ss)} & \colhead{(dd mm ss)} & & & & &  \\
} 
\startdata
1  &  00 39 52.0    &  40 29 44.7      & 50.0$\pm$5.1  &   21.8$\pm$2.2  &                      &                       &                   \\ 
2  &  00 42 20.0    &  41 27 54.7      & 49.3$\pm$4.9  &   23.9$\pm$2.4  & 9.4$\pm$1.0  & 10.4$\pm$1.1  & 4.2$\pm$0.5 \\
3  &  00 42 53.6    &  41 25 51.5      & 46.9$\pm$4.7  &  24.7$\pm$2.5   & 11.4$\pm$1.2 & 14.9$\pm$1.5  & 3.7$\pm$0.4 \\ 
4  &  00 42 55.4    &  41 26 22.8      & 8.1$\pm$0.9  &  4.8$\pm$0.6      & 2.6$\pm$0.3  &  1.7$\pm$0.4  & 0.7$\pm$0.2 \\
5  &  00 43 29.2    &  41 19 02.0      & 11.0$\pm$1.7  &  3.9$\pm$0.7    & 1.5$\pm$0.3  &  1.3$\pm$0.5  & 0.3$\pm$0.2 \\
6  &  00 43 39.1    &  41 26 53.6      & 43.9$\pm$4.4  &  24.6$\pm$2.5  & 10.6$\pm$1.1  & 13.1$\pm$1.4  & 4.0$\pm$0.4 \\
7  &  00 43 54.2    &  41 52 54.7      & 17.1$\pm$1.8   &                         &   3.5$\pm$0.4  & 2.1$\pm$0.3  &                   \\
8  &  00 43 56.4    &  41 47 10.1      & 40.4$\pm$4.2   &                         &   7.0$\pm$0.8  & 5.6$\pm$0.6  & 2.1$\pm$0.3 \\
9  &  00 44 01.0    &  41 21 05.3      & 6.2$\pm$0.8     &  3.6$\pm$0.5   & 0.4$\pm$0.2    & 3.6$\pm$0.7  & 0.4$\pm$0.2 \\
10  &  00 44 05.1    &  41 20 12.7      & 9.9$\pm$1.4   &                        & 3.7$\pm$0.4     &        & 0.9$\pm$0.2 \\
11  &  00 44 26.8    &  41 48 57.4      & 14.2$\pm$2.4  &                        & 3.4$\pm$0.5   &                        &                       \\
12  &  00 44 36.4    &  41 24 57.8      & 86.3$\pm$8.7  &                        & 17.9$\pm$1.8  & 21.7$\pm$2.2  & 9.0$\pm$0.9 \\
13  &  00 44 46.7    &  41 29 23.9      & 38.4$\pm$3.9  &                        & 8.8$\pm$0.9   & 12.6$\pm$1.3  & 4.3$\pm$0.4 \\
14  &  00 44 54.6    &  41 31 52.2      & 76.2$\pm$8.1  &                        & 20.9$\pm$2.2  & 20.3$\pm$2.1  & 5.3$\pm$0.7 \\
15  &  00 45 10.7    &  41 32 53.4      & 22.9$\pm$2.5  &                        & 4.0$\pm$0.5    & 3.8$\pm$0.6  & 3.6$\pm$0.4 \\
16  &  00 45 15.3    &  41 34 27.8      & 14.0$\pm$1.7  &                         & 1.7$\pm$0.4  & 3.6$\pm$0.5  & 3.5$\pm$0.4 \\
17  &  00 45 24.7    &  41 41 00.3      & 22.0$\pm$2.8  &                        & 5.0$\pm$0.6  & 6.4$\pm$0.7  & 1.7$\pm$0.2 \\
18  &  00 45 30.0    &  41 47 36.7      & 20.3$\pm$3.2  &                        & 3.5$\pm$0.5  &                     &                       \\
19  &  00 46 20.2    &  41 53 04.6      & 28.4$\pm$2.9  &                        & 6.9$\pm$0.7  &                     &                   \\
20  &  00 46 30.0    &  41 58 09.4      & 27.9$\pm$2.9  &                        & 5.6$\pm$0.6  &                      &                   \\
\enddata
\end{deluxetable*}

The box method yields the total photon counts in a chosen box, which can be square or rectangular. 
The SNR source counts were measured using different box sizes, generally 4 different sizes from 17 by 17 pixels to 23 by 23 pixels. 
However, more crowded sources would be measured 2 or 3 times using smaller boxes, and some sources required larger boxes due to their unusual shape.  
A few sources were large and irregularly shaped. 
These required a larger box to encapsulate the UV emission, and for these cases small boxes would be used to remove the counts from any stars that fell within the larger box. 
In order to account for background variations, 6 different measurements were done for each filter using manually chosen boxes. 
The average of the 6 measurements was used as the background, and the standard deviation of the 6 measurements used as the uncertainty. 

\begin{figure}[p]
    \centering
    \includegraphics[width = 13.125cm, height = 17.5cm]{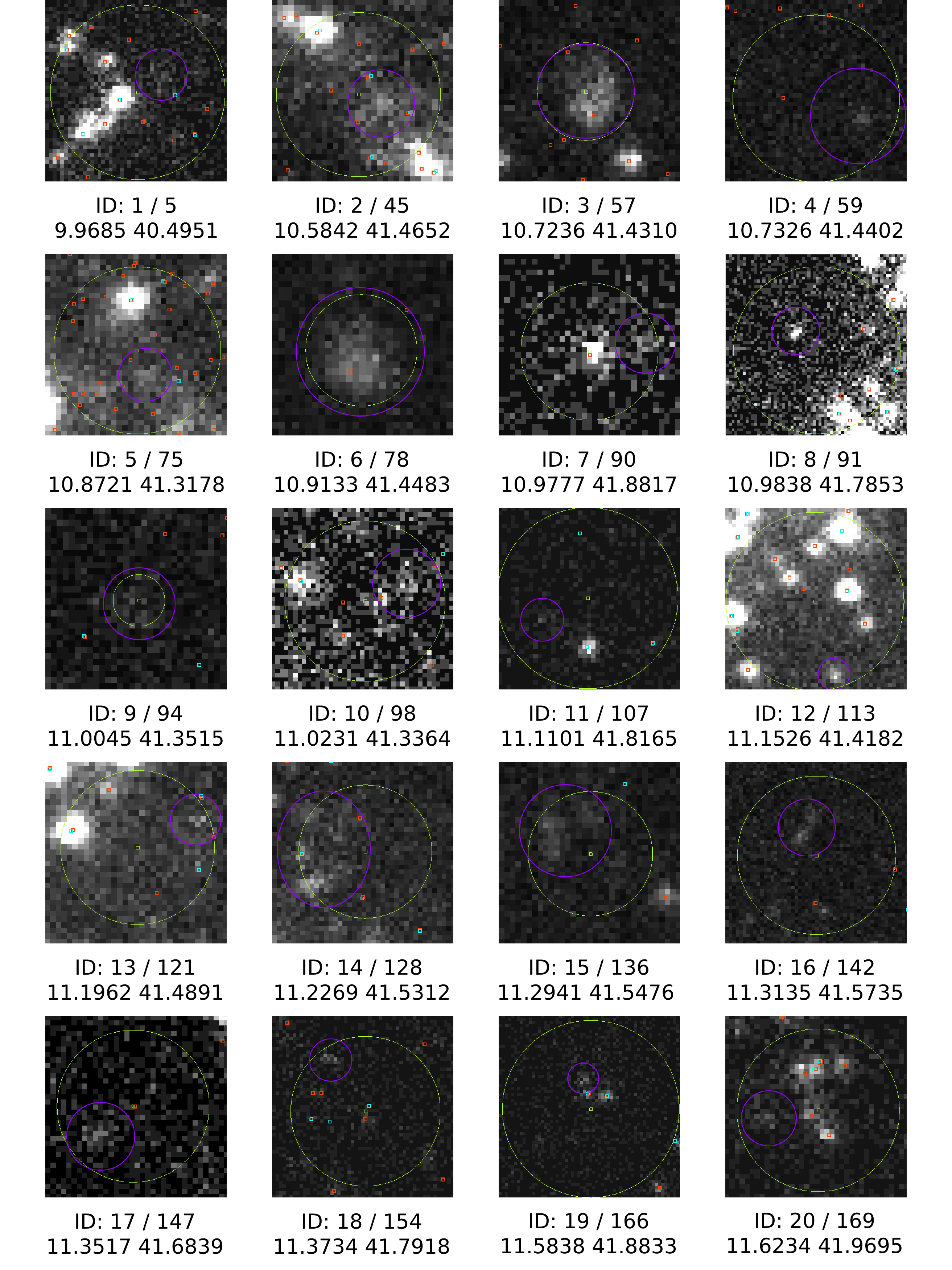}
    \caption{The 20 SNRs with UV emission. SNR ID, Candidate ID and J2000 coordinates are given below each panel.
   The SNR radius is shown by the green circle. 
    The red squares indicate star positions from \citet{2016AJ....152...62M} and \citet{2021AJ....161...79M} , light blue squares indicate positions of stars from GAIA DR3. 
    The purple circles indicate approximately the areas analyzed to obtain source fluxes.} 
    \label{fig:20F148Wimages}
\end{figure}

\begin{figure*}
\gridline{\fig{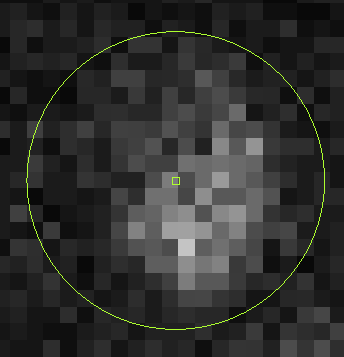}{0.192\textwidth}{ID3(F148W)}
             \fig{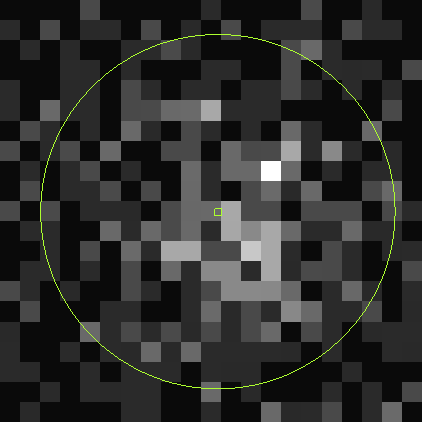}{0.2\textwidth}{ID3(F169M)}
             \fig{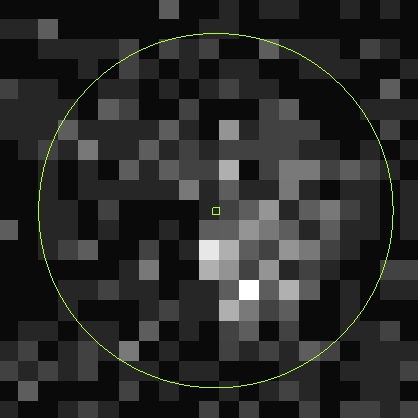}{0.2\textwidth}{ID3(F172M)}
           \fig{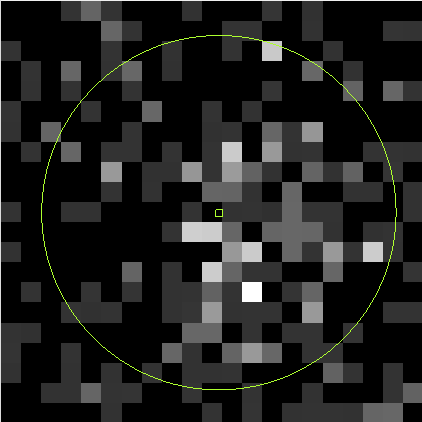}{0.2\textwidth}{ID3(N219M)}
            \fig{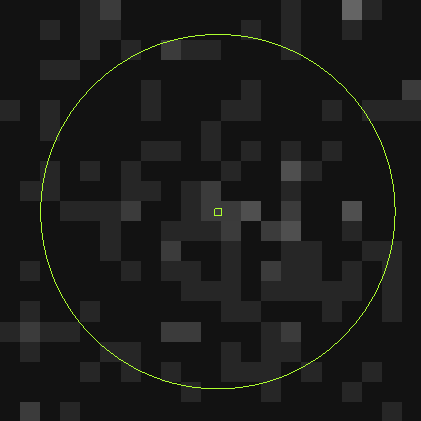}{0.2\textwidth}{ID3(N279N)}
            }
            \gridline{\fig{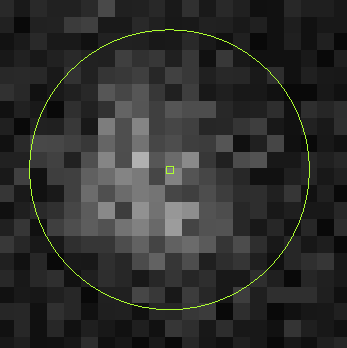}{0.199\textwidth}{ID6(F148W)}
             \fig{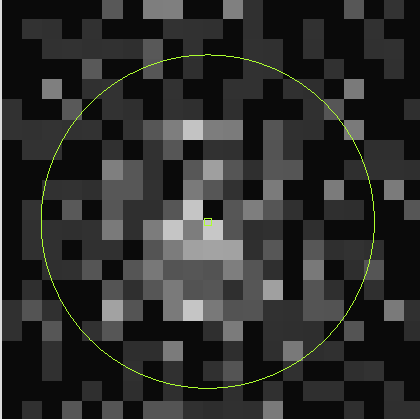}{0.2\textwidth}{ID6(F169M)}
             \fig{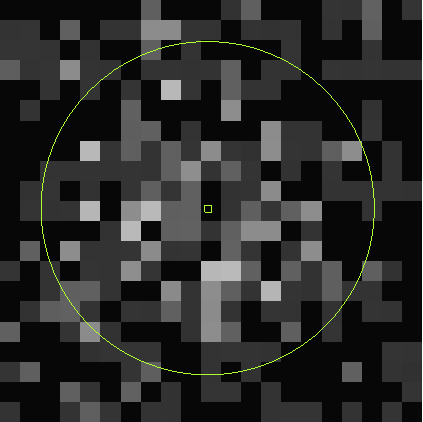}{0.2\textwidth}{ID6(F172M)}
           \fig{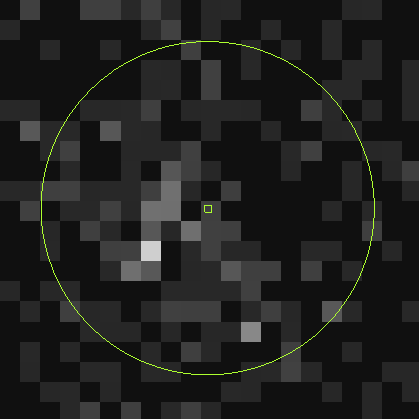}{0.2\textwidth}{ID6(N219M)}
            \fig{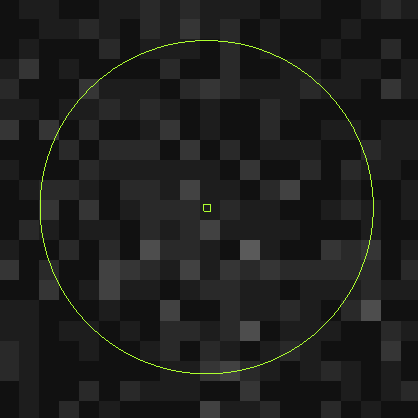}{0.2\textwidth}{ID6(N279N)}
            }
\caption{Images of two of the best selected SNRs in all 5 filters (F148W, F169M, F172M, N219M and N279N, left to right): top row- ID 3; bottom row- ID 6. 
The green circles indicates the SNRs radii.}
 \label{fig:S3S6images}
\end{figure*}

The counts for each filter band image were divided by the net exposure time of each image to obtain a count rate. 
This was corrected for the extended wings of the point-spread-function as given in Table 5 of \citet{2017AJ....154..128T}. 
Then the count rate was converted to flux in units of erg s$^{-1}$ cm$^{-2}$  $\AA^{-1}$ using the flux conversions given in Table 4 of \citet{2017AJ....154..128T}. 
Additional uncertainties include Poisson photon counting errors, uncalibrated spatial variations in detector sensitivity, and uncertainties in correction for the wings of the point spread function.
These were included in quadrature in the calculation of flux errors.
The fluxes and uncertainties were converted into luminosities using the distance to M31 and the effective bandwidth of each filter.
The filter effective wavelengths and effective bandwidths are given in Table 3 of \citet{2017AJ....154..128T}. 

\section{Results and Discussion}  \label{sec:results}

\subsection{UV emission from M31 SNRs and Catalog of UV emitting SNRs in M31} \label{sec:cat}

The images of the 20 M31 SNRs in the F148W band are shown in Figure~\ref{fig:20F148Wimages}. 
These sources exhibit diffuse emission which is not associated with stars, although the strength of the diffuse emission varies.
Images of two of these sources (ID 3 and ID 6) in all five UVIT filter bands are shown in Figure~\ref{fig:S3S6images}.  
The catalog of 20 SNRs in M31 with luminosities in the detected UVIT bands is given in Table~\ref{UVdata}.
Because of some overlap of the fields, some SNRs were imaged in two different fields in the same filter.
The reported band luminosity and uncertainty for those are the average of these measurements. 

\begin{figure*}
\gridline{\fig{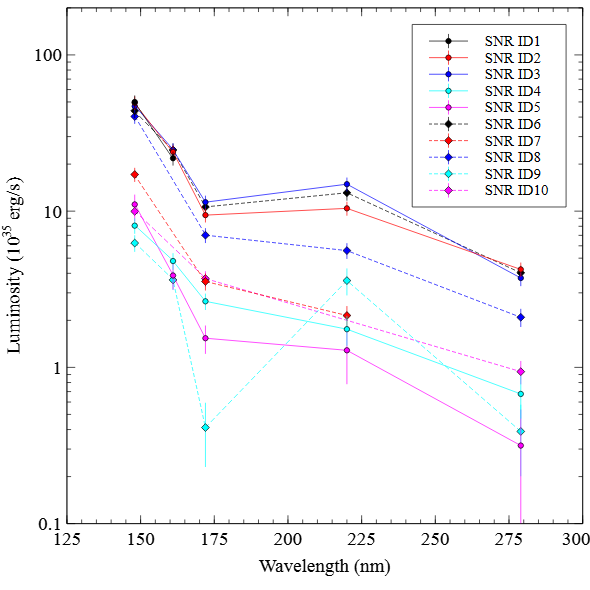}{0.5\textwidth}{(a)}, \fig{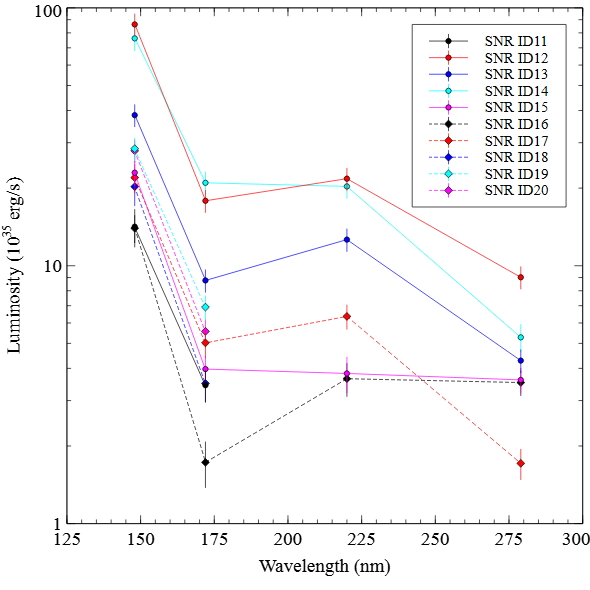}{0.5\textwidth}{(b)}    }
\caption{Spectrophotometry of the 20 SNRs: (a) SNR ID's 1 to 10; (b)SNR ID's 11 to 20. 
\label{fig:20spectra}}
\end{figure*}

The spectral shape of each source is shown in Figure~\ref{fig:20spectra}, panels (a) and (b). 
The F148W filter luminosity is the largest for all SNRs analyzed, in part caused by the wider effective bandwidth for the F148W filter.
The other filter band luminosities vary in prominence, 
however the spectral shapes of the 20 M31 SNRs are similar.

\subsection{Comparison of UV-emitting M31 SNRs with known UV-emitting SNRs}  \label{sec:known}

\begin{deluxetable*}{lllrrrrrrll}
\tablecaption{Ultraviolet emission lines known SNRs (from \citealt{1996ApJS..106..563F})   and the UVIT filter bands which contain those lines.\label{lines}}
\tablewidth{700pt}
\tabletypesize{\scriptsize}
\tablehead{
\colhead{$\lambda$ (A)}            & \colhead{Ion} & \colhead{F148W} & \colhead{F169M} & \colhead{F172M} & \colhead{N219M} & \colhead{N279N} \\
} 
\startdata
 {1334.53}          & C II                    & X    &          &        &        &       \\
 {1355.6}           & O I                     & X    &          &        &        &       \\
 {1371.29}          & O V                     & X    &          &        &        &       \\
 {1393.76}          & Si IV                   & X    &          &        &        &       \\
 {1397.20, 1399.77} & O IV{]}                 & X    &          &        &        &       \\
 {1398.13, 1404.77} & S IV{]}                 & X    &          &        &        &       \\
 {1402.77}          & Si IV                   & X    &          &        &        &       \\
 {1483.32, 1486.50} & {[}N IV{]},N IV{]}      & X    & X        &        &        &       \\
 {1533.43}          & Si II                   & X    & X        &        &        &       \\
 {1548.20, 1550.77} & C IV                    & X    & X        &        &        &       \\
 {1574.8}           & {[}Ne V{]}              & X    & X        &        &        &       \\
 {1601, 1602}       & {[}Ne IV{]}             & X    & X        &        &        &       \\
 {1640 blend}       & He II                   & X    & X        & X      &        &       \\
 {1660.81, 1666.15} & O III{]}                & X    & X        & X      &        &       \\
 {1670.81}          & Al II                   & X    & X        & X      &        &       \\
 {1730 blend}       & N III{]}                & X    & X        & X      &        &       \\
 {1746.82, 1748.61} & N III{]}                & X    & X        & X      &        &       \\
 {2320.95, 2331.40} & {[}O III{]}             &      &          &        & X      &       \\
 {2323.50, 2324.69} & C II{]}                 &      &          &        & X      &       \\
 {2328.51, 2334.40} & Si II{]}                &      &          &        & X      &       \\
 {2795.52, 2802.70} & Mg II                   &      &          &        &        & X    \\
\enddata
\end{deluxetable*}

\begin{deluxetable*}{lllrrrrrrll}
\tablecaption{Known UV-emitting SNRs and estimated UVIT filter band luminosities calculated from measured line fluxes\label{knownSNRs}}
\tablewidth{700pt}
\tabletypesize{\scriptsize}
\tablehead{
            &      &           &               & \multicolumn{5}{c}{Estimated UVIT Filter Band Luminosity (erg/s)}               &                    &        \\
            \colhead{SNR}        & \colhead{Type} & \colhead{Location}  & \colhead{Distance} & \colhead{F148W}     & \colhead{F169M} & \colhead{F172M}   & \colhead{N219M}   & \colhead{N279N}    & \colhead{Aperture Size} & \colhead{Source} \\
        &  &  & \colhead{(pc)} &      &  &    &    &     &  & 
} 
\startdata
            &      &           &               & 5.82E+36 & 4.48E+36 & 2.29E+36 & 1.53E+36 & -        & 10''x20''          & a      \\
Cygnus Loop & CC   & Milky Way & 725           & 3.62E+36 & 2.77E+36 & 8.46E+35 & 1.17E+36 & 2.10E+35 & 4 x 10''x20''      & b      \\
            &      &           &               & 5.18E+36 & 4.12E+36 & 2.28E+36 & 1.39E+36 & -        & 3 x 3.8''x12.4''   & c      \\
Vela        & CC   & Milky Way & 250           & 3.56E+35 & 2.53E+35 & 8.24E+34 & 6.07E+34 & -        & 4 x 10''x20''      & b      \\
PuppisA     & CC   & Milky Way & 2,146         & 1.86E+33 & 1.43E+33 & 2.88E+32 & -        & -        & 10''x56''          & d     \\
N49         & CC   & LMC       & 50,000        & 5.47E+37 & 4.94E+37 & 8.40E+36 & 2.60E+37 & 1.30E+37 & 5 x 10''x20''      & e      \\
N103B       & Ia   & LMC       & 45,990        & 4.62E+35 & 3.19E+35 & 5.06E+34 & -        & -        & 2.5'' (circular)   & f  \\
N132D       & CC   & LMC       & 52,000        & 4.34E+34 & 2.51E+34 & 8.07E+33 & 6.21E+33 & 8.49E+33 & 3 x 1'' (circular) & g      \\
E0102-7219  & CC   & SMC       & 64,386        & 6.28E+33 & 3.45E+33 & 1.09E+33 & 5.28E+32 & 9.24E+32 & 1''x1''            & g  \\
            &      &           &               & 4.05E+35 & 2.12E+35 & 3.18E+34 & 1.50E+34 & 1.92E+34 & 10''x20''          & h   \\   
\enddata
\tablecomments{a) \cite{1980ApJ...238..881R}. b) \cite{1981ApJ...246..100R}. c) \cite{1988ApJ...324..869R}.  d) \cite{1995ApJ...454L..35B}. e) \cite{1992ApJ...394..158V}.  f) \cite{2020ApJ...902..153B}. g)  \cite{2000ApJ...537..667B}. h) \cite{1989ApJ...338..812B}. }
\end{deluxetable*}

The spectra of known UV-emitting SNRs shows that their emission in the UVIT filter bands would be dominated by emission lines, with a small contribution from continuum radiation. 
A list of UV emission lines measured in known SNRs \citep{1996ApJS..106..563F} is presented in Table~\ref{lines}, together
with the UVIT filter bands that contain these lines. 
These SNRs include 3 Galactic SNRs- the Cygnus Loop \citep{1980ApJ...238..881R,1981ApJ...246..100R,1988ApJ...324..869R}, 
Vela \citep{1981ApJ...246..100R}, and Puppis A \citep{1995ApJ...454L..35B};
3 SNRs in the Large Magellanic Cloud- N132D \citep{2000ApJ...537..667B},  
N49  \citep{1992ApJ...394..158V}, and N103B \citep{2020ApJ...902..153B};
and 1 SNR in the Small Magellanic Cloud- E0102-7219 \citep{1989ApJ...338..812B,2000ApJ...537..667B}.
Table~\ref{knownSNRs} lists these SNRs, with SNR Type, distance, and aperture size for the line flux measurements given in the references.

\begin{figure*}
\gridline{\fig{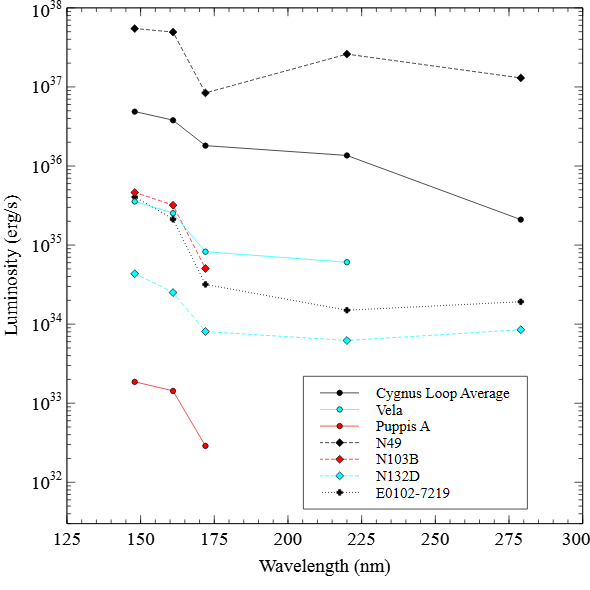}{0.5\textwidth}{(a)}, \fig{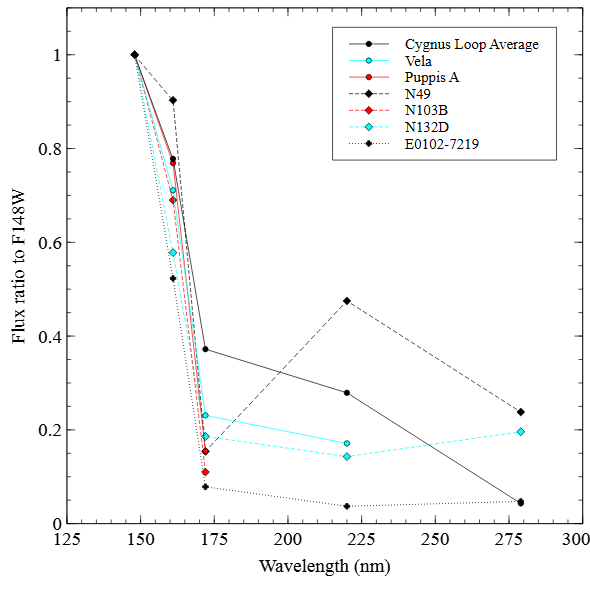}{0.5\textwidth}{(b)}    }
\gridline{\fig{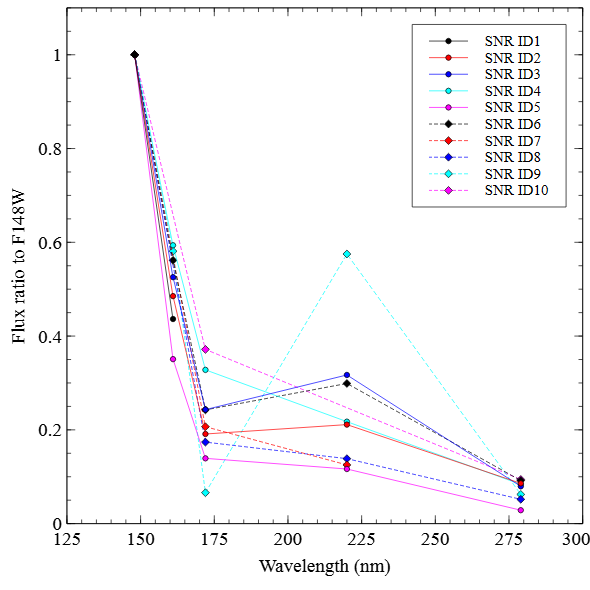}{0.5\textwidth}{(c)}, \fig{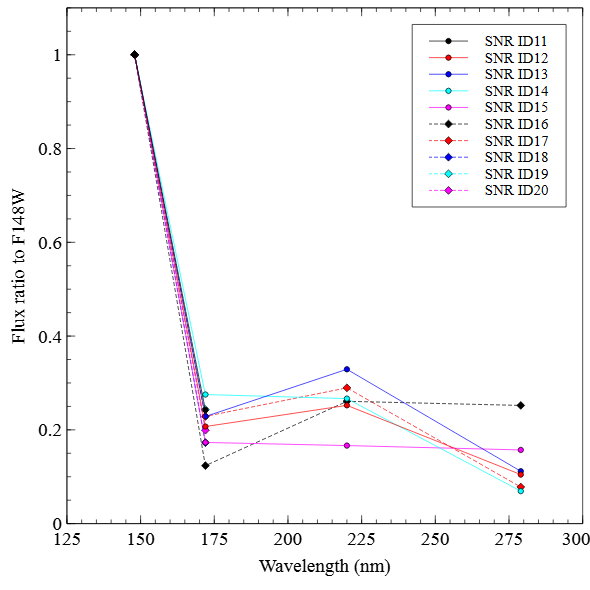}{0.5\textwidth}{(d)}    }
\caption{Spectral shape of 7 known UV-Emitting SNRs: (a) estimated band luminosities (for Cygnus Loop the average of the 3 values and for E0102-7219 the second listed values from Table~\ref{knownSNRs} are plotted); (b) scaled band luminosities (luminosity divided by the F148W luminosity of each SNR). 
Scaled band luminosities for the SNRs in M31: (c) SNR ID's 1 to 10; (d) SNR ID's 11 to 20.
\label{fig:rescaled}}
\end{figure*}

We estimated the UVIT band luminosities for the known SNRs by summing the published luminosities or fluxes (converted to luminosities) of the detected emission lines which fall within each UVIT filter, as listed in Table~\ref{lines}.
The results of this are given in columns 5 through 9 of Table~\ref{knownSNRs} and the band luminosities vs. the effective wavelengths of the UVIT filter bands are shown in panel (a) of Figure~\ref{fig:rescaled}.
Although the band luminosities of the different sources are quite variable, the F148W luminosity is the brightest.
That band includes more emission lines than the other bands (see Table~\ref{lines}). 

Supernovae and SNRs can be categorized into two separate types: thermonuclear runaway (Type Ia) and core collapse (CC). 
The small sample of known SNRs with UV spectroscopy includes only one type Ia SNR (N103B).
This SNR has band luminosities (panel (a) of Figure~\ref{fig:rescaled}) which does not appear different to the CC-type SNRs.
Scaled band luminosities are defined as the band luminosities for each SNR divided by the F148W band luminosity for that SNR.
The scale band luminosities of the seven known UV-emitting SNRs are shown in panel (b) of Figure~\ref{fig:rescaled}.
They are all quite similar, with factor $\sim$2 variations between the different SNRs.

Although UV spectroscopic data for the M31 SNRs does not exist, we can compare the UV emission of the M31 sources to known UV-emitting SNRs by comparing the UVIT band luminosities. 
For several of the known 
SNRs, the aperture only covered part of the area of the SNR, so that we do not have a good measurement 
of the line luminosities over the whole SNR. 
The extrapolation from the aperature to the whole SNR is a highly uncertain factor because of line brightness variations over the face of the SNR. 
However, if the line ratios are nearly constant over the face of the SNR, the scale band luminosities should be representative of the spectral shapes of the whole SNR.
Thus we compare the spectral shapes of the known SNRs to one another and to the M31 SNRs using the scaled band luminosities.

For the band luminosities, known SNRs (panel (a) of Figure~\ref{fig:rescaled}) show larger variations than the 20 new UV emitting SNRs in M31 (panels (a) and (b) of Figure~\ref{fig:20spectra}). 
This is, in part, caused by the different fractions of the SNR area measured in the aperture of the spectroscopic observations (Table~\ref{lines}).  
However, the scaled band luminosities (panels (b) of Figure~\ref{fig:rescaled}) show that the spectral shapes of the known SNRs are quite similar. 
The F148W value is the largest, followed by the F169M value, then with similar values in the remaining 3 bands (F172M, N219M and N279N).
The  scaled band luminosities of the 20 M31 SNRs (panels (c) and (d) of Figure~\ref{fig:rescaled}) have remarkably similar variations to the known SNRs.
The strong similarity of the rescaled band luminosities of the newly detected 20 M31 SNRs to known UV-emitting SNRs (Figure~\ref{fig:rescaled}) is evidence that the detected UV emission is from the SNR, rather than from foreground or background objects.

This current list of 20 UV-emitting SNRs in M31 is likely incomplete. 
The interstellar extinction of UV light from  M31 by the Milky Way's ISM is small, but the ISM of M31 can have significant extinction. 
The extinction has been measured for stellar clusters in M31 using UVIT photometry of stellar clusters \citep{2022AJ....164..183L}:
the mean extinction is $E(B-V)$=0.24 and ranges from 0 to 0.6.
Thus SNRs within or on the farside of M31's disc may be undetected because of extinction. 
More than half (Section~\ref{sec:selection}) of the SNRs from the original list of 177 were excluded because of confusion with stellar emission: there was not enough diffuse emission separated from overlapping stars within the SNR radius for detection. 

\subsection{Physical conditions of the 6 M31 SNRs with X-ray spectra}  \label{sec:SNRmodel}

\begin{deluxetable*}{llllllllllll}
\tablecaption{Models for 6 M31 SNRs with X-ray spectra\label{xraySNRs}}
\tablewidth{700pt}
\tabletypesize{\scriptsize}
\tablehead{
            \colhead{SNR$^{(a)}$}        & \colhead{Type$^{(b)}$} &  \colhead{$M_{ej}^{c)}$} & \colhead{(s,n)$^{(d)}$} & \colhead{Shock$^{(e)}$}  & \colhead{Age} & \colhead{E0}     & \colhead{density} & \colhead{$\dot{M}/(4\pi V_{w})$}   & \colhead{$kT_{2}^{(e)}$}   & \colhead{$EM_{2}^{(e)}$} & \colhead{Consistent?$^{(f)}$}   \\ 
 SPH11 ID   &    &  \colhead{$M_{\odot}$} & & & \colhead{(yr)} &  \colhead{($10^{51}$erg)}    & \colhead{(cm$^{-3}$)}  &    \colhead{(gm/cm)}  &  \colhead{(keV)}   &  \colhead{($10^{58}$cm$^{-3}$)}    
} 
\startdata
1066(UV)&	Ia & 1.2	&   (0,7) & fwd  &1970	&	12.1	&	0.33	&	 &	20.5		& 5.76$\times 10^{-3}$& Y \\
1066(UV)&	Ia & 1.2	&   (2,12) & fwd  &324	&	33.2	&	& 9.39$\times 10^{13}$ 	&	0.70	&	48.4 & N \\
1066(UV)&	Ia & 1.2	&   (2,12) & rev  & 133	&	16.0	& &	3.66$\times 10^{13}$	&	20.6	&	1.12 & Y\\
\hline
1275(UV) & Ia & 1.2	&   (0,7) & fwd  &	14040	&0.311&	0.594	&&	1.243&	8.23$\times 10^{-3}$ & Y\\
1275(UV) & Ia & 1.2	&   (2,12) & fwd &2950		&0.387	&&1.50$\times 10^{14}$	&	0.035&	131 & N  \\
1275(UV) & Ia & 1.2	&   (2,12) & rev     &636		&6.76		&&5.85$\times 10^{13}$	&	1.03	&	3.05  & Y\\
\hline
      1055      &  (unk.)    &    5       &    (0,7) & fwd  &    7220	&	0.234	& 0.080 &	&	0.131 &	0.421  & Y   \\
      1055      &  (unk.)    &    5       &    (2,12) & fwd &  1420	&	0.769	& & 8.65$\times 10^{13}$	&	0.035 &	96.6   & N  \\
      1055      &  (unk.)    &    5       &    (2,12) & rev &  292	&	14.8	& &	3.37$\times 10^{13}$ &		1.03	&	2.24   & Y \\
\hline
1234	&	CC & 10	&   (0,7) & fwd 	&23300	&1.02	&	0.488	& &	0.683 &	9.33$\times 10^{-2}$& Y \\
1234	&	CC & 10	&   (2,12) & fwd  &5400	&	2.03	&&	3.18$\times 10^{14}$	&	0.033	& 368 & N \\
1234	&	CC & 10	&   (2,12) & rev & 1030	&	38.8	&&	6.35$\times 10^{13}$	&	0.974	 & 2.24  & Y\\
\hline
1535&	CC & 10	&   (0,7) & fwd 	&	18500 &	0.344	&0.320	&&	0.254&	8.19$\times 10^{-2}$ & Y\\
1535&	CC & 10	&   (2,12) & fwd 	&3360		&2.49		&&1.30$\times 10^{14}$	&	0.035	&78.8 & N \\
1535&	CC & 10	&   (2,12) & rev	&782		&37.3		&&5.07$\times 10^{13}$	&	1.03	&	1.83 & Y\\
\hline
1599	&	CC & 10	&   (0,7) & fwd 	&18200	&1.143&	0.568	&	&0.774 &	1.120 & Y\\
1599	&	CC & 10	&   (2,12) & fwd 	&3640		&3.45		&&2.98$\times 10^{14}$			&0.044&	361 & N \\
1599	&	CC & 10	&   (2,12) & rev  	&807		&56.9	&&	1.16$\times 10^{14}$		&1.299	&8.37  & Y\\
\enddata
\tablecomments{(a)The UV-emitting SNRs are marked with (UV). (b) Type Ia (Ia), core collapse (CC) or unknown (unk.). (c) Ejecta mass taken as 1.2M$_{\odot}$ for Ia, 10M$_{\odot}$ for CC or 5M$_{\odot}$ for unk . (d) s=power law index for circumstellar medium density: constant s=0, or wind s=2;   n=power law index for ejecta density.  (e) Measured $kT$ and $EM$ are assumed to be from forward shock (fwd) or from reverse shock (rev); if "fwd" then $kT_2$ and $EM_2$ are the predicted values for the reverse shock; if "rev" then $kT_2$ and $EM_2$  are the predicted values for the forward shock. (f) Is the predicted $EM_2$ small enough to be consistent with observations? (Y=yes), (N=no).}
\end{deluxetable*}

To determine density of environment and evolution status of SNRs, X-ray observations of the thermally emitting shocked gas are required. 
There are 6 SNRs in M31 which have had their X-ray spectra analyzed with hot plasma models from shocked gas, from \citet{2012AA...544A.144S} and \citet{2018AA...620A..28S}.
We use the spectral parameters for  SPH11 SNRs 1050 and  1066 from \citet{2012AA...544A.144S}  and for SPH11 SNRs 1234, 1275, 1535 and 1599 from \citet{2018AA...620A..28S}\footnote{SPH11 1234 was analyzed in both: we used the newer  analysis from the later reference.}.
The emission measures ($EM$) for each SNR was calculated from "norm" or from the flux, if no "norm", using XSPEC and the best fit spectral model. 

The modelling software used is SNRpy \citep{2019AJ....158..149L,2017AJ....153..239L} which is based on the unified models of SNR evolution of \citet{1999ApJS..120..299T}, with extensions added, including non-equilibrium ionization.
For each SNR, we fit the measured shocked-gas temperature $kT$ and emission measure $EM$ for 3 different cases:
i) explosion in a uniform environment (s=0) and emission from the forward-shocked gas; ii) explosion in a stellar wind (s=2) and emission from the forward-shocked gas; and iii) explosion in a stellar wind (s=2) and emission from the reverse-shocked gas. 
Explosions in a uniform medium lead to much brighter emission from forward-shocked gas, but explosions in a wind environment can lead to either brighter emission from the forward-shocked gas or from the reverse-shocked gas. 
The results of the models are given in Table~\ref{xraySNRs}.
Because the measured X-ray spectrum is for the brighter component (i.e. forward or reverse-shocked), we mark the models which
are consistent with the observations (i.e. measured $EM$ larger than $EM_2$) in the Table. 
Emission from the forward-shock gas in a uniform medium (s=0) or emission from the reverse-shocked gas in a stellar wind  (s=2) are consistent with observations.
Other information can help to select a preferred model for each SNR. 
E.g., SNRs with energy above $10^{52}$ erg should be rare \citep{2022ApJ...931...20L,2020ApJS..248...16L,2017ApJ...837...36L}. This disfavors the (s,n)=(2,12) reverse-shock models for SPH11 1234, 1535 and 1599. 
The energies, ages and densities (or stellar wind parameters) are reasonable for the other models. 

The 2 UV-emitting SNRs with X-ray spectra, SPH11 1066 and 1275, are marked with (UV) in Table~\ref{xraySNRs}.
The clear difference between the UV emitting SNRs and the UV non-detected SNRs is that the 2 UV-emitting SNRs are Type Ia.
Because of the large age of Type Ia progenitors compared to CC progenitors, Type Ia are expected to be at significant height above the disk plane, and thus to have low extinction.
The models for the 6 SNRs have a range of ages, explosion energies and densities, and the sample is too small to see systematic differences in the ages, densities or explosion energies between the UV-detected and non-detected SNRs.  
One would expect the column density (given in \citealt{2012AA...544A.144S} and \citealt{2018AA...620A..28S}) to be smaller for the UV-detected SNRs.
The column densities are small, with large uncertainties, except for SPH11 1599 which is not detected in UV. 
Better determination of column densities with better X-ray spectra are needed to confirm the relation between UV-detection and low column density.
 
\subsection{Comparison of numbers of M31 SNRs in different wavebands}  \label{sec:SNRnumbers}

\begin{figure}[h]
\centering
\gridline{\fig{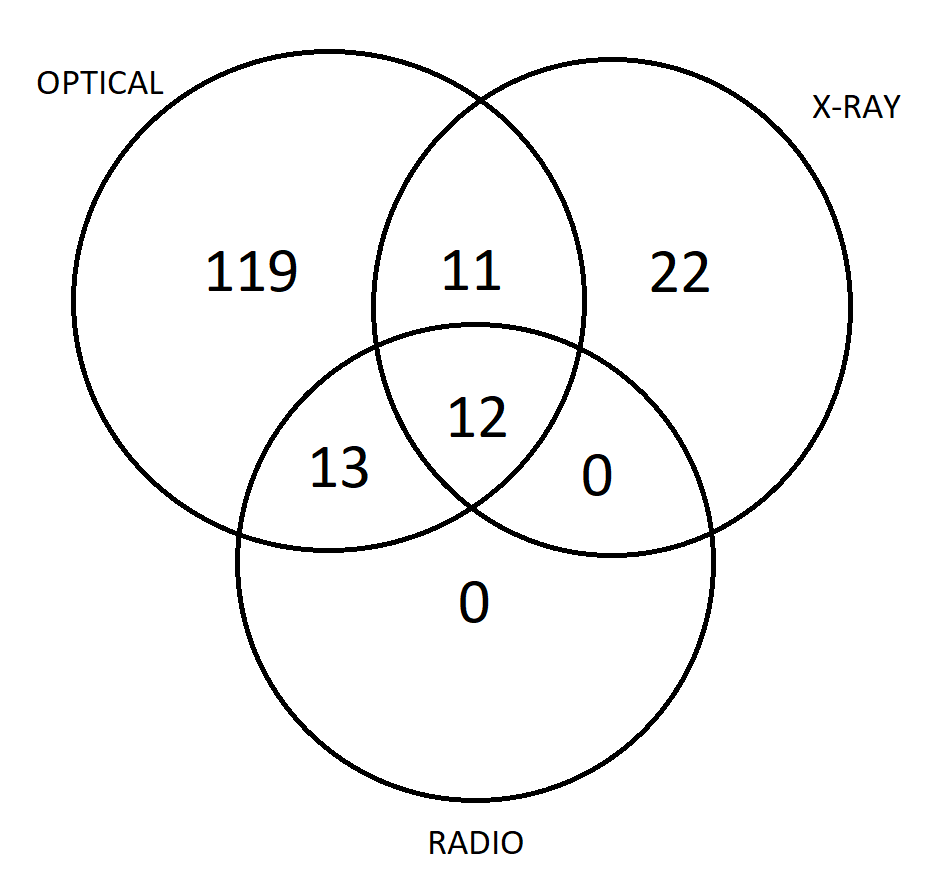}{0.5\textwidth}{(a)}
             \fig{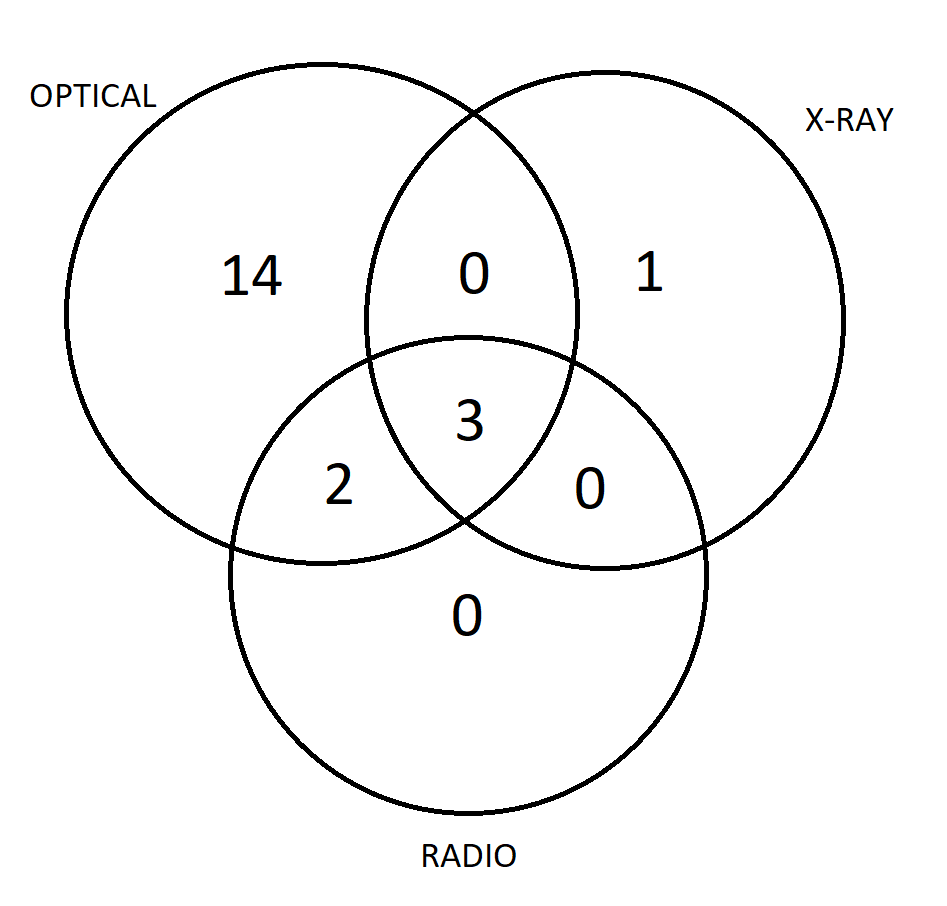}{0.5\textwidth}{(b)}
            }
    \caption{Venn diagrams for SNRs in M31 detected in different wavebands (optical, X-ray and radio: (a): all SNRs; (b): for the 20 SNRs detected in UV (this work).} 
    \label{fig:venn}
\end{figure}

The 177 SNRs in M31 were detected in different wavebands; optical, X-ray and radio, prior to the current work.
Panel (a) of Figure~\ref{fig:venn} shows the numbers of SNRs detected in single and multiple wavebands using a Venn diagram.
\citet{2000ApJ...544..780P} shows similar Venn diagrams (optical, radio, and x-ray detected SNRs) for the nearby galaxies M33 and NGC300. M33 has a total number of 109 SNRs and SNR candidates and NGC300 has a total of 44, compared to the total of 177 SNRs and SNR candidats for M31. 
For all 3 galaxies (M31, M33 and NGC300), optically-detected SNRs dominate with fractions of detected SNRs of  155/177, 79/109 and 28/44, respectively. 
The X-ray and radio detections are  significantly smaller fractions: the X-ray detected fractions are  45/177, 21/109, and 6/44 respectively; and the radio detected fractions are 25/177, 53/109 and 17/44, respectively. 
The optically detected fractions are similar ($\sim0.6-0.9$) for all 3 galaxies; the radio detected fractions higher for M33 and NGC300 ($\sim0.4-0.5$) than for M31 ($\sim0.15$), and the X-ray detected fraction  higher for M31 ($\sim0.25$) than for M33 and NGC300 ($\sim0.15-0.2$).
This is probably the result of the intensive X-ray observations of M31 \citep{2012AA...544A.144S,2018AA...620A..28S} compared to the other two galaxies.
The three sets of optically-detected, radio-detected and X-ray detected SNRs show little overlap for all three galaxies. 
This is consistent with the opposing selection effects for these 3 wavebands, as discussed in \citet{2000ApJ...544..780P}:
SNRs identified through optical represent those located in regions with relatively low confusion from H$\alpha$ emission, well away from star-forming regions;  
radio-selected SNRs are biased toward star-forming regions; and 
X-ray SNRs are selected for soft X-ray spectra and association with H II regions, so are biased against SNRs with hard spectra and no optical counterparts.

\citet{2017ApJS..230....2B} considers the statistics of optical, radio and X-ray detections for the Large Magellanic Cloud (LMC). 
The Venn diagram of SNRs for the LMC shows most SNRs (47 of 59) are detected in all three bands.
This is probably a result of sensitivity to luminosity because of the closer distance of the LMC ($\sim$20 times closer than the other nearby galaxies discussed above). 
The Venn diagrams given in \citet{2017ApJS..230....2B} include NGC300 and M33, with similar numbers to those given in \citet{2000ApJ...544..780P}, and M31 with results similar to Figure~\ref{fig:venn} here.
The other galaxies with Venn diagrams in  \citet{2017ApJS..230....2B} are the Small Magellanic Cloud (SMC), NGC7793, NGC 6946, NGC 55, and one diagram for NGCs 2403, 3077, 4214, 4395,  4449 and 5204 combined (hereafter referred to as "NGCcombined").
For the SMC, which is also nearby, the diagram is similar to that for the LMC with most SNRs detected in all 3 bands.
The diagrams for NGCcombined, NGCC 7793 and NGC 6946 are dominated by SNRs detected in optical only, like M31, M33 and NGC 300.
NGC 55 has too few SNRs (6 total) to draw conclusions on numbers. 
The differences are likely the result of the differing sensitivities of the observations in the different wavebands for each of the galaxies, as discussed by \citet{2017ApJS..230....2B}. 

  The 20 UV-emitting SNRs in M31 are listed in Table~\ref{tab:optXray}. 
The other wavebands in which these 20 SNRs are detected are as follows.  
Sources 1, 2, 4, 5, 8, 10, 11, 12, 14, 16, 17, 18, 19, and 20 were listed only in \citet{2014ApJ...786..130L}  (optical); 
sources 7 and 9 were listed in \citet{2014ApJ...786..130L} and in \cite{1993A&AS...98..327B}(optical \& radio); 
source 15 was listed only in \citet{2012AA...544A.144S}  (X-ray) and
sources 3, 6, and 13 were listed in all 3 (optical, X-ray, \& radio).
Panel (b) of Figure~\ref{fig:venn} shows the numbers of UV-emitting SNRs detected in optical, X-ray and radio.
Nearly all of the UV-emitting SNRs (19 of 20) are detected in optical. 
This is  not surprising because the emission mechanisms for UV and optical are most similar: forbidden and recombination lines from shock-ionized gas. 
In contrast, the radio emission mechanism is synchrotron from shock-accelerated electrons and the the emission mechanism is primarily thermal bremmstrahlung with some contribution from lines.

The total number of UV-emitting SNRs in M31 is similar to the number radio emitting SNRs (20 vs. 25) but much smaller than numbers of optical or X-ray emitting SNRs.
The fractions of UV-emitting SNRs in M31 are: 19/155 (optical), 5/25 (radio) and 4/45 (X-ray). 
These have not been measured for other galaxies yet, but are similar ($\sim0.1-0.2$) for the 3 different categories. 
This indicates that the UV-selection criterion is different than the optical, radio and X-ray criteria listed above.
The most important expected UV detection criterion is source extinction. 
This is important for the 7 previously known SNRs, which all have low extinction (see references in Table~\ref{knownSNRs}).
The UV extinction to SNRs in M31 is dominated by line-of-sight distance through M31's disk\footnote{The Milky Way contribution to extinction is small in the direction of M31}. 
For M31 SNRs detected in  optical, radio and X-ray, the disk extinction is small at those wavelengths so they should be detected independent of distance into the disk. 
The SNRs detected in UV will be those on the near side of the disk.
The fraction detected in UV to other wavebands should be determined by extinction, i.e. disk geometry, and thus approximately constant and equal to the fraction of disk which is on the near side of M31 and with low extinction in UV.  
The visible to UV extinction curve \citep{2007ApJ...663..320F} has E($\lambda$-V)/E(B-V) $\sim$3.5 to 7 (average $\sim$5) for $\lambda=$120 nm to 280 nm (with the peak at 220 nm). 
Typical measured E(B-V) values in M31 are in the range 0.0 to 0.6 \citep{2022AJ....164..183L}, which 
yields E($\lambda$-V) from 0 to 3 mag, and extinction factors from 1 to 0.06. 
The fraction of UV to optical SNRs in M31 (19/155$=0.12$, Figure~\ref{fig:venn}) is consistent with that expected from extinction, but the differing sensitivity of  UV and optical observations could also affect the fraction.

\section{Conclusion}  \label{sec:conclusion}

Using the survey images of M31 carried out by Astrosat's UVIT, we searched for diffuse UV-emission from M31 SNRs. 
SNRs for analysis were obtained from previous 
optical, X-ray and radio surveys for SNRs in M31.
We used stellar catalogues and the UV images to remove SNRs contaminated with stellar emission, enabling us to detect 20 SNRs with diffuse UV emission in M31. 
Band fluxes for the five observed UVIT filters, F148W, F169M, F172M, N219M and N279N, were measured for these 20 SNRs to obtain band luminosities. 
The result is the first catalog of UV emitting SNRs in M31.

The band luminosities of the UV-emitting SNRs were compared to the band luminosities computed form the spectra for seven previously known UV-emitting SNRs in the Milky Way, the LMC, and the SMC. 
We find similar spectral shapes between the known SNRs and the M31 SNRs.
The spectral shapes and the diffuse nature of the emission together form good evidence that the UV emission from the
20 M31 SNRs is dominated by line emission, like known SNRs, and that the UV emission is associated with the SNRs.

For the small sample of 6 SNRs in M31 with X-ray spectral models, we apply SNR models to obtain their physical characteristics. 
The 2 UV-emitting X-ray SNRs are Type Ia, the other 4 X-ray SNRs are CC type.  
Type Ia indicates positions above the disk plane in M31. 
The two UV-emitting X-ray SNRs have low measured extinction in X-rays, so that it is consistent that the detection of UV emission is related to low extinction.
 
We compare the numbers of SNRs detected in M31 for different wavebands (optical, radio and X-ray) to 
those detected in other nearby galaxies given by \citet{2000ApJ...544..780P} and  \citet{2017ApJS..230....2B}. 
This confirms the somewhat opposing selection effects for detecting SNRs in the different wavebands, as discussed by \citet{2000ApJ...544..780P}. 
19 of the 20 UV-emitting SNRs are detected in optical, which is expected because the emission mechanisms for both UV and optical are forbidden and recombination lines from shock-ionized gas.

It is desirable to carry out spectroscopic observations to confirm the line nature of the UV emission from these SNRs, although spectroscopy will difficult for the typically crowded regions in M31 where the SNRs are located.

\begin{acknowledgments}
This work was supported by a grant from the Canadian Space Agency. 
The authors thank the reviewer for making a number of suggestions to improve this manuscript.
\end{acknowledgments}


\end{document}